\documentclass[aps,pra,showpacs,twocolumn,amsmath,amssymb,superscriptaddress,nofootinbib]{revtex4}
\usepackage{graphicx}
\usepackage{bm}
\usepackage{amssymb}
\usepackage{amsmath}
\usepackage{latexsym}
\usepackage{color}
\newcommand{\vc}[1]{\mbox{\boldmath $#1$}} 
\newcommand{\ind}[1]{_{#1}}    
\newcommand{\indrm}[1]{_{\mathrm {#1}}}    
\newcommand{\gammah}{\gamma_{\ind{H}}}   
\newcommand{\gammao}{\gamma_{\ind{0}}}   
\begin{document}  
\title{Output coupling from x-ray  free-electron laser cavities with intracavity beam splitters}
\author{Yuri Shvyd'ko} \email{shvydko@aps.anl.gov}
\affiliation{Advanced Photon Source, Argonne National Laboratory,  Argonne, Illinois 60439, USA}
\begin{abstract} 
  Permeable mirrors are typically used for coupling photons out of
  laser cavities.  A similar approach was proposed for output coupling
  photons from the cavities of x-ray free-electron laser (XFEL)
  oscillators. One of the Bragg-reflecting crystal mirrors is thin,
  just a few extinction length, and is used as a permeable mirror. 
  However, this method is very often limited to extractions of only a
  few tenths of the intracavity power. Other cavity-based XFELs, such
  as the high-gain regenerative amplifier XFEL, require much higher
  outcoupling efficiency. Here, alternative schemes are proposed and
  analyzed for coupling x-ray photons out of XFEL cavities using
  intracavity Bragg-reflecting, x-ray-transparent diamond crystal beam
  splitters.  The intracavity beam splitters are efficient and
  flexible in terms of the amount of the power they are capable of
  coupling out of the cavity, an amount that can be varied promptly
  from zero to close to 100\%.  The schemes can be readily extended to
  multi-beam outcoupling.
\end{abstract}
\date{\today}

%
%

\maketitle

\section{Introduction}
The next generation of high-repetition-rate x-ray free-electron lasers
(XFELs) will allow for optical cavity 
feedback, like in classical lasers. Unlike self-amplified spontaneous
emission (SASE) XFELs \cite{KS80,BPN84,EAA10}, cavity-based XFELs
\cite{HR06,KSR08} are capable of generating fully coherent x-ray beams
of high brilliance and stability.  Two major optical cavity-based XFEL
schemes are presently under discussion: low-gain and high-gain.

\begin{figure}[t!]
\includegraphics[width=0.5\textwidth]{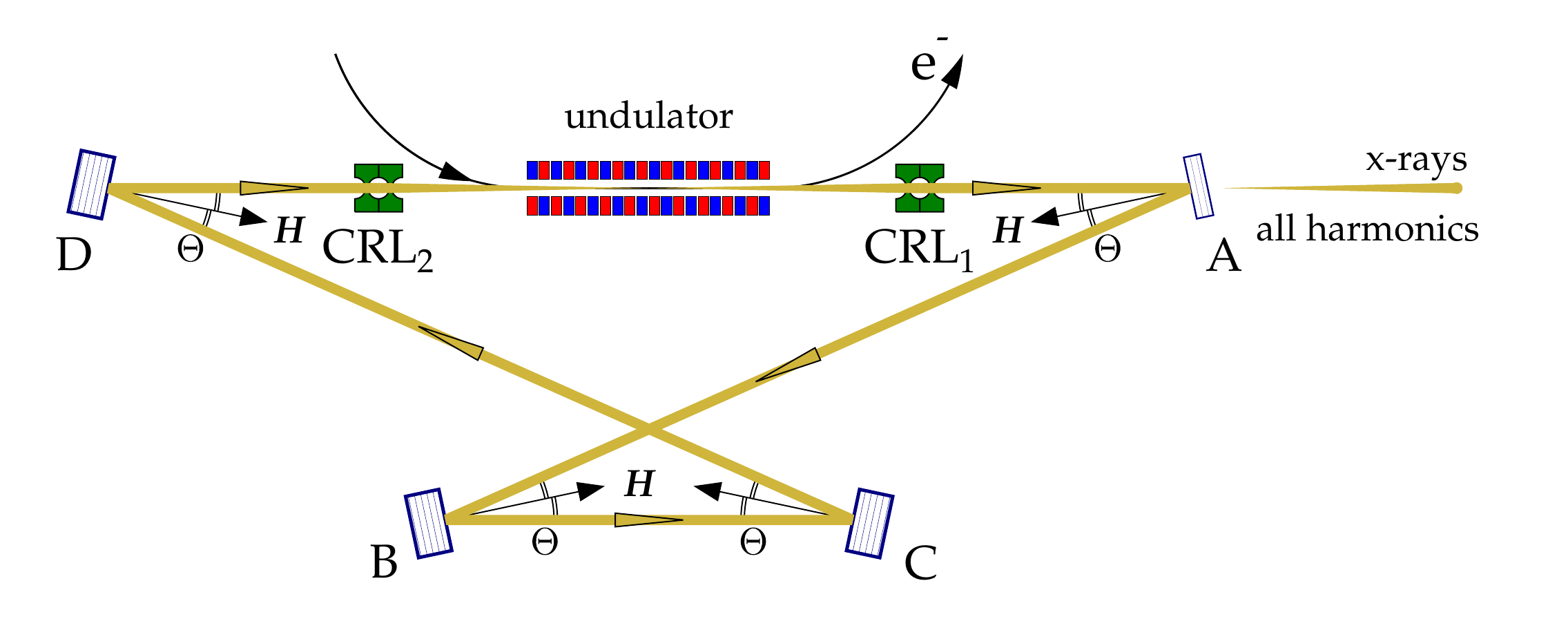}
\caption{Schematic of a cavity-based XFEL with a tunable optical
  cavity composed of four Bragg-reflecting flat-crystal mirrors
  (A,B,C, and D) and compound refractive lenses (CRLs) as focusing
  elements \cite{KS09}.  X-ray power is coupled out of the cavity
  though permeable thin  crystal A with reduced reflectivity. The
  incidence and reflection angles $\Theta=\pi/2-\theta$ are the same
  for all crystals, where $\theta$ is Bragg's angle.}
\label{fig001}
\end{figure}

An x-ray free-electron laser oscillator (XFELO)
\cite{KSR08,KS09,LSKF11} is a low-gain cavity-based XFEL, which
requires a low-loss (high-Q) cavity.  XFELOs are promising to generate
radiation of unprecedented spectral purity (a few-meV bandwidths).
Figure~\ref{fig001}(a) shows an example schematic of the XFELO with a
tunable cavity composed of four Bragg-reflecting flat-crystal
``mirrors'' (A,B,C, and D) and compound refractive lenses (CRLs) as
collimating and focusing elements \cite{KS09}.  Alternative cavity
designs can be considered as well, such as a tunable compact
non-coplanar six-crystal cavity \cite{Shv13} or others.  X-rays
generated by electrons in the undulator circulate in the low-loss
optical cavity composed of flat Bragg-reflecting diamond crystals
with close to 100\% reflectivity \cite{SSB11} and low-absorbing Be
paraboloidal lenses  \cite{LST99,KSG18} stabilizing the cavity.  

Because XFELOs are low-gain devices, the outcoupling efficiency is
typically required to be a few percent.  A standard procedure for
outcoupling in laser physics by using a partially reflective mirror
can be extended to a hard x-ray regime, as well. Indeed, a similar
approach was proposed for coupling photons out of the XFELO optical
cavities \cite{KSR08,KS09,Shv13} by using thin, permeable crystal
mirrors.



Another possible realization of the cavity-based XFEL is a high-gain
regenerative amplifier FEL (XRAFEL).  It was first demonstrated in the
infrared \cite{NSF99} and is also considered in the hard x-ray regime
\cite{HR06,MDD17,FSS19}. The XRAFEL optical cavity can be either the
same tunable four-crystal cavity shown in Fig.~\ref{fig001}, or a
six-crystal cavity \cite{Shv13}, or any alternative one. The XRAFEL is
a high-gain FEL, which can reach saturation after a few round-trip
passes. It can therefore allow for a high (close to 100\%) extraction
efficiency.



How  can this be achieved?  Can the permeable thin-crystal approach be
extended to high-efficiency outcoupling required for XRAFEL? Are there
other options?  High-efficient output coupling using Bragg-reflecting
pinhole crystal mirrors was proposed and studied in \cite{FSS19}.
Here an alternative possibility is discussed, using intracavity,
Bragg-reflecting, x-ray-transparent diamond crystal beam splitters.


\begin{figure}[t!]
\includegraphics[width=0.5\textwidth]{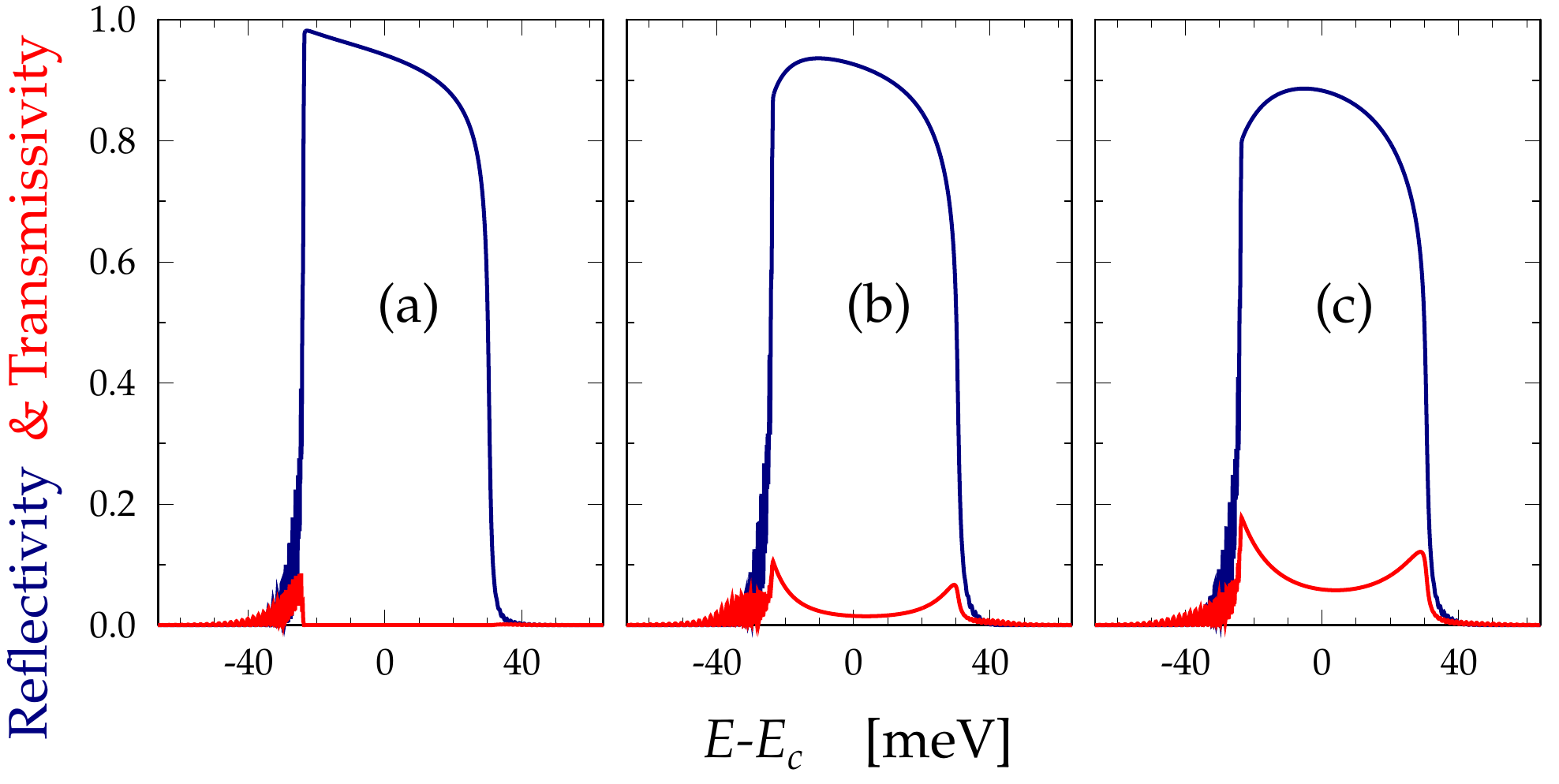}
\caption{Spectral reflection profiles (blue) of the four-crystal perfectly
  aligned {\em empty} cavity (see Fig.~\ref{fig001}) as a result of four
  successive (400) Bragg reflections of x-rays from diamond crystals
  B$\rightarrow$C$\rightarrow$D$\rightarrow$A.  The transmission
  profile through crystal A is shown in red. Crystals B, C, and D have
  a thickness of 300~$\mu$m, while crystal A's thickness is (a)
  300~$\mu$m, (b) 20~$\mu$m, and (c) 15~$\mu$m, respectively. Photon
  energy $E_{\indrm{c}}=6.9558$~keV, angular spread (FWHM) of x-rays -
  $2.5~\mu$rad, Bragg's angle $\theta=88^{\circ}$
  ($\Theta=2^{\circ}$). The spectral reflection width $\Delta
  E_{\ind{4\times (400)}}\simeq 54$~meV.
}
\label{fig4x400}
\end{figure}


\section{Permeable crystal output coupling}
\label{thincrystal}

The required for the XFEL cavity crystals with close to 100\% Bragg
reflectivity of x-rays (98\%-99.5\%) can be achieved only if x-ray
transparent crystals are used, for which the ratio of the extinction
length $\bar{\Lambda}_{\ind{H}}$ to the photo-absorption length
$L_{\indrm{a}}$ is very small, $\bar{\Lambda}_{\ind{H}}/L_{\indrm{a}}
\simeq 1/100$, as in diamond \cite{SSB11}. The extinction
length\footnote{Here we are using the extinction length as defined in
  \cite{SL12}. In other texts, e.g. in \cite{Authier} an alternative
  definition is used
  $\Lambda_{\ind{H}}=2\pi\,\bar{\Lambda}_{\ind{H}}$.}
\begin{align}
\bar{\Lambda}_{\ind{H}}\,=\,& \frac{\sqrt{\gammao|\gammah|}}{\sin\theta} \bar{\Lambda}_{\ind{H}}^{(s)}, \label{eq0010}  \\
  \gammao\,=\,  \sin(\theta+\eta), & \hspace{0.5cm} \gammah=\sin(\eta-\theta)
\label{eq0020} 
\end{align}
is a measure of penetration of x-rays into the crystal in Bragg
diffraction with diffraction vector $\vc{H}$. Here, $\theta$ is the
glancing angle of incidence to the reflecting atomic planes (Bragg's
angle), $\eta$ is the asymmetry angle between the crystal entrance
surface and the reflecting atomic planes, and
$\bar{\Lambda}_{\ind{H}}^{(s)}$ is the extinction length in the
symmetric scattering geometry, when
$\eta=0$. $\bar{\Lambda}_{\ind{H}}^{(s)}$ is to a good accuracy a
Bragg reflection invariant in low-absorbing crystals like diamond. The
$\bar{\Lambda}_{\ind{H}}^{(s)}$ values for the allowed Bragg
reflections in diamond can be found tabulated in \cite{SL12}.

Reflectivity $R(0)$ and transmissivity $T(0)$ of  a thick ($d >
\bar{\Lambda}_{\ind{H}}$) x-ray-transparent (low-absorbing) crystal at
the center of the Bragg reflection region are
\begin{align}
R(0)\simeq & 1- 4\exp\left(-{d}/{\bar{\Lambda}_{\ind{H}}} \right),\\
T(0)\simeq & 4\exp\left(-{d}/{\bar{\Lambda}_{\ind{H}}}  -{d}/{L_{\indrm{a}}\gammao}\right),
\end{align}
see, for example Eq.(7.11) of \cite{Authier}.
Therefore,  in addition to x-ray transparency, the high reflectivity requires that the
crystal thickness is $d \gtrsim 10 \bar{\Lambda}_{\ind{H}}$
\cite{Shvydko-SB}.

To ensure low losses in the cavity, almost all crystals should have a
very high reflectivity and therefore should be sufficiently
thick. These are crystals B, C, and D in the example shown in
Fig.~\ref{fig001}. The crystals are in symmetric scattering geometry
($\eta=0$) to avoid angular dispersion \cite{Shvydko-SB,SLK06}, which
can deteriorate the transverse profile of the beam.\footnote{A lateral
  shift of $\simeq \bar{\Lambda}_{\ind{H}}\cos\theta $ of x-rays also
  takes place in the symmetric scattering geometry \cite{LS12,SL12}.}
However, one crystal may have reduced reflectivity and therefore some
transmissivity to allow for a certain portion of the intracavity beam
to be coupled out of the cavity.

Because $\bar{\Lambda}_{\ind{H}}/L_{\indrm{a}}\ll 1$, to ensure, say,
approximately a $5$\% transmission, the crystal thickness should be
reduced to $\simeq 4.4 \bar{\Lambda}_{\ind{H}}$.  Typically
$\bar{\Lambda}_{\ind{H}}\simeq 4-10~\mu$m for Bragg reflections in
diamond crystals used to backreflect photons with energies in the
$7-10$-keV range \cite{SL12}. Therefore, the thickness of a diamond
crystal with the 5\%-outcoupling efficiency should be about
$15-45~\mu$m.

Figure~\ref{fig4x400} shows examples of spectral Bragg reflection
profiles (blue lines) upon successive Bragg reflections from diamond
crystals in the four-crystal cavity with the crystals set into the
$\vc{H}$=(4 0 0) Bragg reflection. The transmission spectra through
thin crystal A are shown in red. The profiles were calculated by using
the dynamical theory of x-ray Bragg diffraction with crystals B, C, and D
being 300~$\mu$m thick, and output coupler crystal A being 300~$\mu$m
(a), 20~$\mu$m (b), 15~$\mu$m (c). In the latter case the outcoupling
efficiency is close to about 5\%, and the ratio of the crystal thickness
$d=15~\mu$m to the extinction length $\bar{\Lambda}_{511}=3.6~\mu$m is
$d/\bar{\Lambda}_{511}\simeq 4.2$. Such thin diamond crystals can be
manufactured and handled without degrading Bragg diffraction
performance \cite{KVT16}. However, extracting more power from the
cavity would require much thinner crystals, which are both very
difficult to manufacture without introducing crystal defects and very
difficult to handle without degrading Bragg diffraction performance.

An even bigger challenge occurs if Bragg reflections with the smallest
diffraction vectors $\vc{H}$ for which $\bar{\Lambda}_{\ind{H}} \simeq
1-2~\mu$m have to be used to backreflect x-ray photons with energies
below 5~keV. In all of these cases the thickness required even for the
5\%-outcoupling efficiency crystal becomes extremely small, $d\simeq
4.4\bar{\Lambda}_{\ind{H}} \simeq 5-10~\mu$m.

Altogether, outcoupling through a permeable thin crystal is typically
limited to a less than $\simeq 5$\% efficiency, unless high-indexed
Bragg reflections with $\bar{\Lambda}_{\ind{H}} \gtrsim 15~\mu$m are
used that are appropriate for handling x-rays with photon energies $E
\gtrsim 15$~keV.

\section{Beam-splitter output coupling}
\label{beamsplitter}

\begin{figure}[t!]
\includegraphics[width=0.5\textwidth]{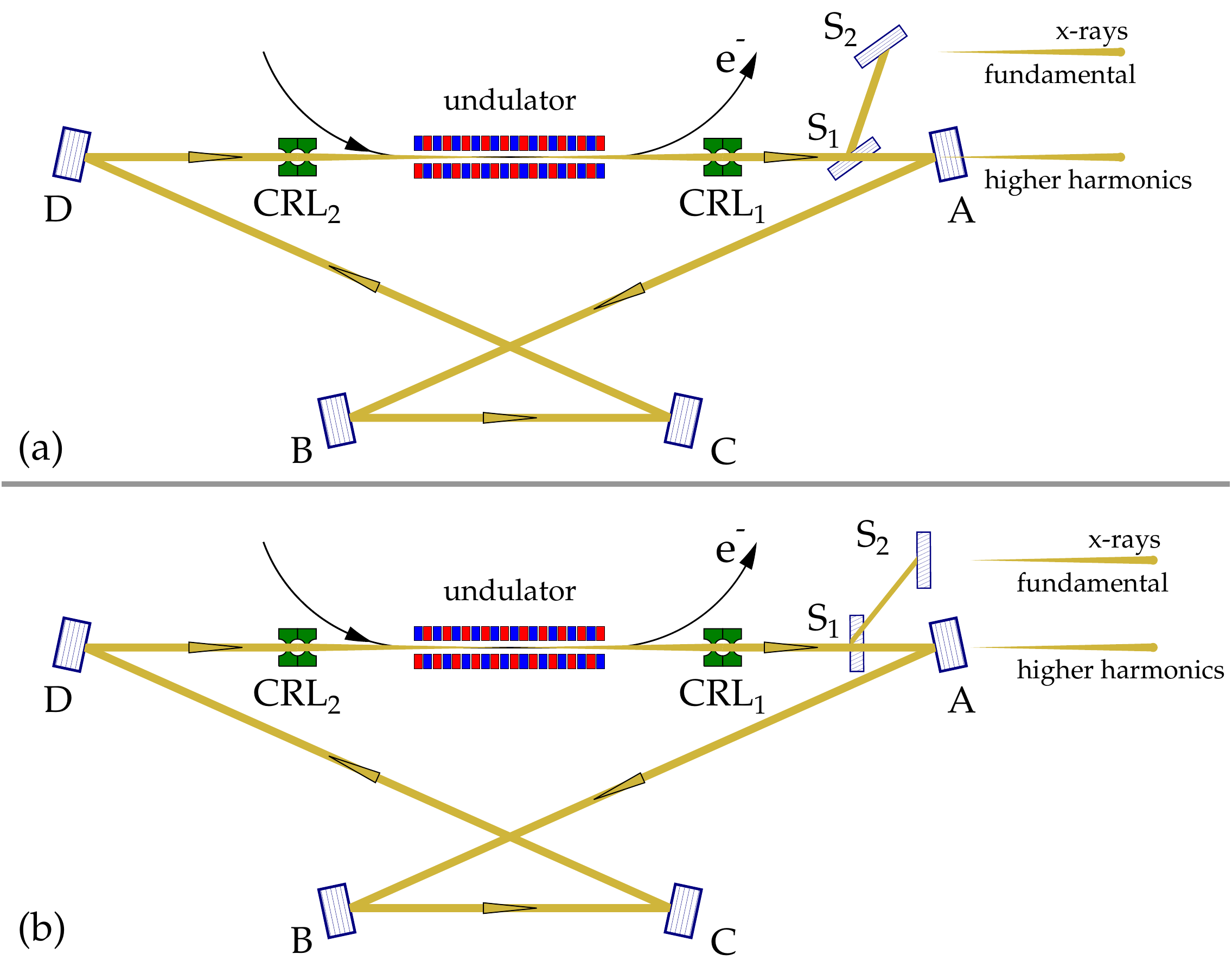}
\caption{``Beam-splitter'' output coupling methods presented in the
  example of the four-crystal cavity. In contrast to the scheme of
  Fig.~\ref{fig001}, all four Bragg-reflecting crystals (A, B, C, and D)
  are thick and are featuring highest reflectivity for the fundamental
  harmonic.  The fundamental is now coupled  out through a thin
  beam-splitter crystal S$_1$, while higher harmonics can be coupled
  out through transparent-for-them crystal A. Beam-splitter crystal
  S$_1$ (output coupler) and crystal S$_2$ designed for fixed-exit
  outcoupling can be either in the reflection (Bragg-case) geometry (a) or
  in transmission (Laue-case) geometry (b).  Other types of beam splitters
  can be used instead of the Bragg-reflecting crystals as well.}
\label{fig001LB}
\end{figure}

Here we present an alternative method of extracting intracavity
radiation power from XFEL cavities, which may have a much higher than
5\% efficiency, and in particular may be appropriate for XRAFEL
optical cavities. Again, we use  the four-crystal cavity as an example,
although the technique is applicable to other cavity types as well.

In this approach, all crystals including A are thick ($d \gtrsim 10
\bar{\Lambda}_{\ind{H}}$) and are featuring high-reflectivity for the
fundamental harmonic; see cavity schematics in Fig.~\ref{fig001LB}.
The fundamental is outcoupled through an additional beam-splitter
crystal S$_1$ inserted into the intracavity beam and set into Bragg
diffraction either in the reflection (Bragg-case) scattering geometry
as in Fig.~\ref{fig001LB}(a), or in the transmission
(Laue-case) scattering geometry 
as in Fig.~\ref{fig001LB}(b). 
The higher harmonics can be still outcoupled through crystal A, whose
thickness could be tailored to be transparent for them. To minimize
losses in the cavity, the thickness of the beam-splitter crystal
should be chosen to be much smaller than the absorption length. This
requirement is, however, much less demanding than the
a-few-extinction-lengths requirement for the output coupler crystal in
the ``permeable crystal'' method discussed in
Sec.~\ref{thincrystal}. The amount of outcoupled intracavity power is
changed by varying the crystal reflectivity, which can be achieved
either by changing crystal angle or crystal thickness, as discussed
below.

There is a second crystal S$_{\ind{2}}$ with the identical Bragg
reflection in the non-dispersive ($+-$) setting, which is used to
direct the out-coupled radiation parallel to the undulator axis
independent of the photon energy. Use of two additional crystals, with
all four in the ($+--+$) configuration, could be also considered to
return the x-ray beam to the undulator axis.

Diamond crystals as the beam splitters would be the preferred choice,
because of their high x-ray transparency, resilience to radiation and
heat load, and high Bragg reflectivity. In fact, diamond crystals have
been in use for a long time as Bragg-reflecting beam splitters for
multiplexing x-ray beams at storage ring \cite{AFG94,MSG98} and XFEL
\cite{ZFS14,STB14,FAB15} facilities. In those applications, they are
used to extract a narrowband component from the primary broadband
beam. In the present case, the task is to extract a certain amount of
the intracavity power without changing its spectral composition.


\begin{figure}[t!]
\includegraphics[width=0.5\textwidth]{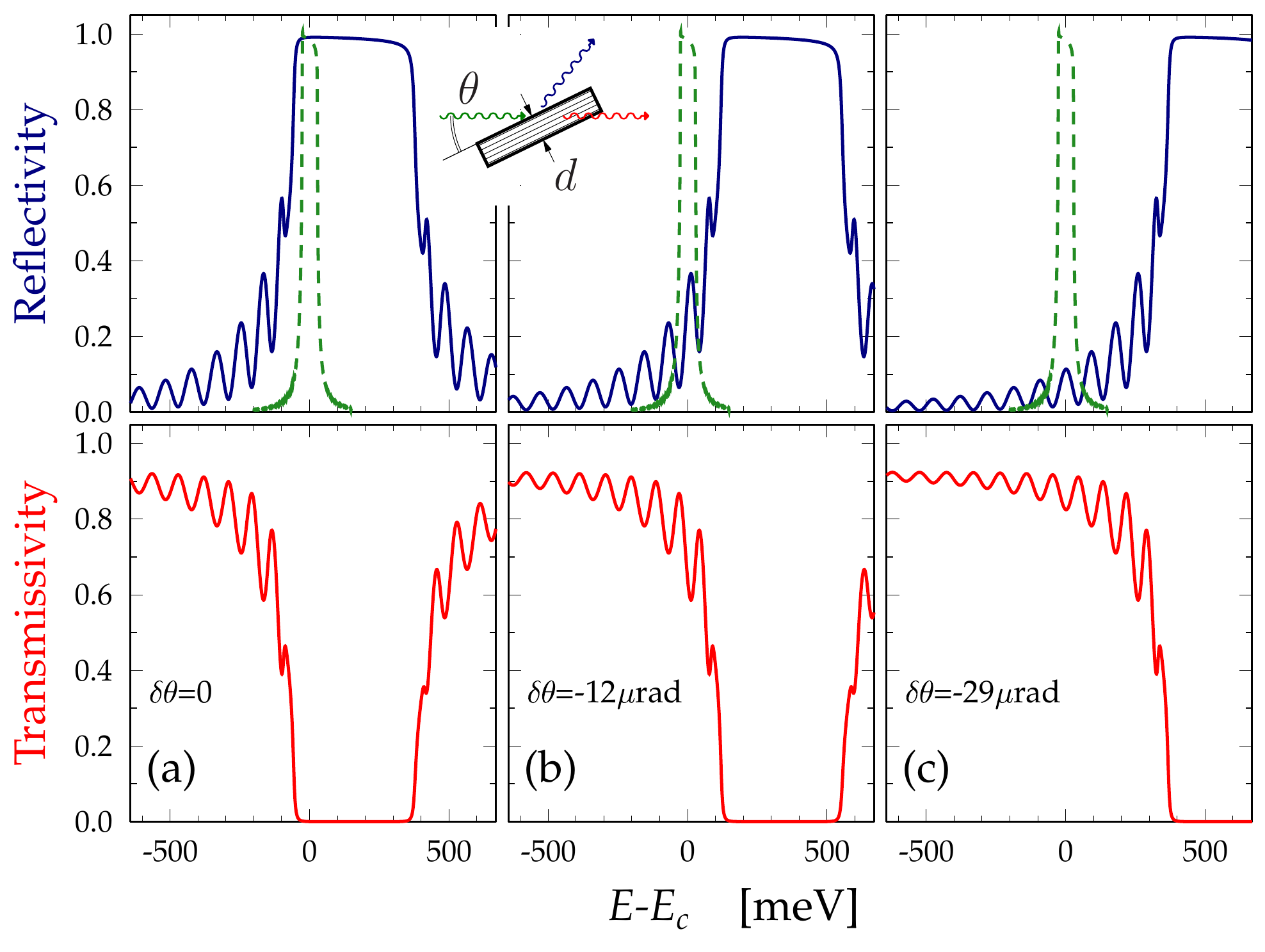}
\caption{Reflection (blue) and transmission (red) spectral profiles of
  x-rays from a $d=15$-$\mu$m-thick diamond beam-splitter crystal
  $S_{\ind{1}}$ in the symmetric (111) Bragg diffraction
  ($\eta=0$). The x-rays are at the incidence angle
  $\theta=\tilde{\theta}+\delta\theta$
  ($\tilde{\theta}=25.6434^{\circ}$) to the diffraction planes (111):
  (a) $\delta\theta=0$; (b) $\delta\theta=-12~\mu$rad; (c)
  $\delta\theta=-29~\mu$rad. The green dashed line is a reference
  spectral Bragg reflection profile from 300-$\mu$m-thick diamond
  crystals in the cavity (as in Fig.~\ref{fig001LB})
  which are set to the (400) reflection with Bragg's angle
  $88^{\circ}$ and centered at $E_{\ind{c}}=6.9558$~keV.}
\label{fig003}
\end{figure}
\begin{figure*}[t!]
\includegraphics[width=0.99\textwidth]{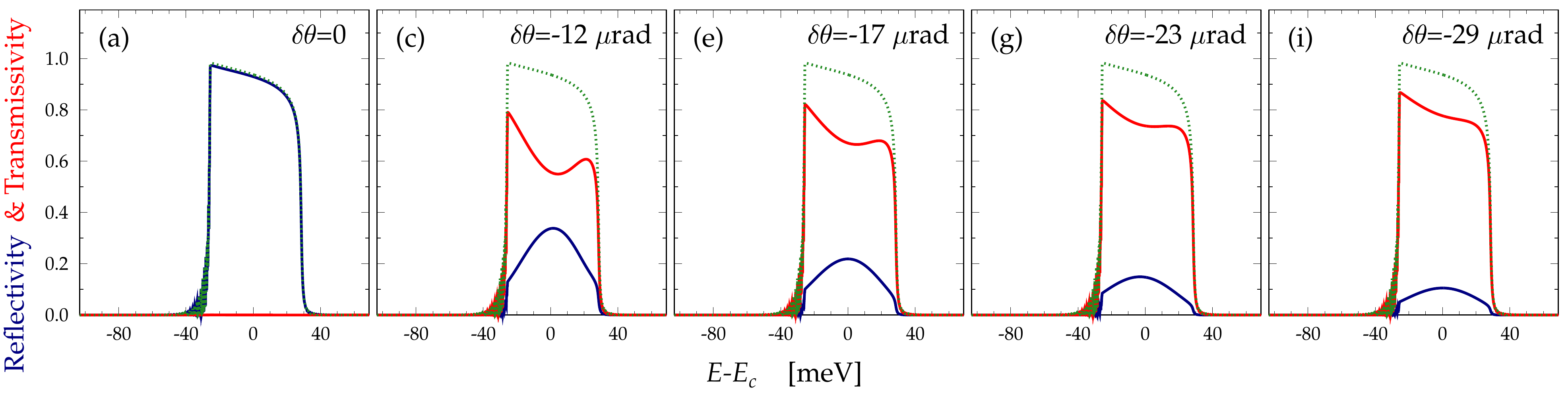}
\caption{Spectral profiles: (dotted green) of the radiation in the
  diamond four-crystal cavity upon successive (400) Bragg reflection
  from 300-$\mu$m-thick diamond crystals
  A$\rightarrow$B$\rightarrow$C$\rightarrow$D with the central cavity
  energy $E_{\indrm{c}}=6.9558$~keV; (blue) of the radiation coupled
  out of the cavity with a beam-splitter crystal in the Bragg-case
  geometry; (red) of the radiation transmitted through the beam
  splitter and left in the cavity.  The beam splitter is a
  15-$\mu$m-thick diamond crystal set in the (111) Bragg reflection at
  incidence angles $\theta=\tilde{\theta}+\delta\theta$
  ($\tilde{\theta}=25.6434^{\circ}$) as in Fig.~\ref{fig003}.}
\label{fig002}
\end{figure*}

\subsection{Bragg-case beam splitter}

We consider first the Bragg-case beam splitter as in
Fig.~\ref{fig001LB}(a), and assume in all following examples (unless
specified otherwise) that the cavity crystals are set into the (400)
Bragg reflection with the Bragg angle of $88^{\circ}$. This sets the
center of the Bragg reflection range at
$E_{\indrm{c}}=6.9558$~keV.

The spectral Bragg reflection profile from a thick crystal ($d \gg
\bar{\Lambda}_{\ind{H}}$) in the reflection (Bragg-case) scattering
geometry features close to a 100\% reflectivity in a spectral range
$\Delta E_{\ind{H}} = E\ (d_{\ind{H}}/\pi \bar{\Lambda}_{\ind{H}})$
around the photon energy $E=hc/\lambda$ defined by Bragg's law
$\lambda =2d_{\ind{H}}\sin\theta$.  Here, $d_{\ind{H}}$ is the
interplanar distance of the reflecting atomic planes corresponding to
the diffraction vector $\vc{H}$, and $\theta$ is the glancing angle of
incidence to the reflecting atomic planes (Bragg's angle).
Figure~\ref{fig003}(a) shows an example of the Bragg reflection
profile (solid blue line) from diamond in the (111) reflection.  The
initial Bragg angle $\theta=25.6434^{\circ}$ is chosen such that the
high-reflectivity range overlaps with the cavity spectral profile
(shown by dashed green line) centered at $E_{\indrm{c}}=6.9558$~keV).
The solid red line in the bottom graph shows the corresponding spectral
transmission dependence. The (111) Bragg reflection parameters are:
$d_{\ind{111}}=2.059$~\AA, $\bar{\Lambda}_{\ind{111}}=1.1~\mu$m, and
$\Delta E_{\ind{111}}=520$~meV.

\begin{figure}[t!]
\includegraphics[width=0.5\textwidth]{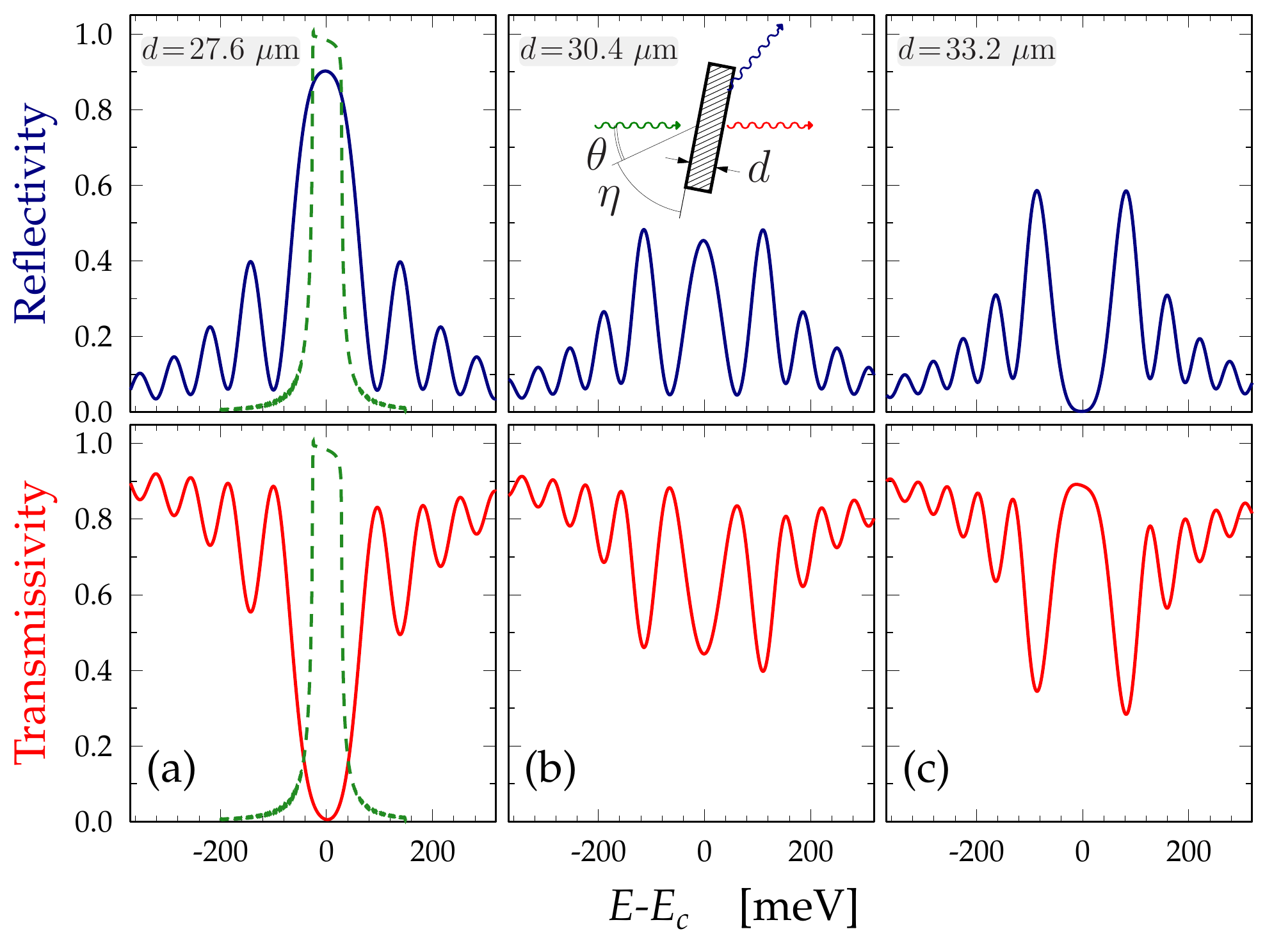}
\caption{Reflection (blue) and transmission (red) spectral profiles of
  x-rays in a diamond beam-splitter crystal $S_{\ind{1}}$ set to the
  (111) Bragg diffraction in Laue-case geometry with
  $\eta=54.74^{\circ}$ 
[the entrance crystal surface is parallel to
  the (100) planes]. The profiles are calculated for the crystal
  thickness $d$ varying with an increment of $\Lambda_{\ind{H}}/4$ of the Pendel\"osung
  period $\Lambda_{\ind{H}}=11~\mu$m: (a) $d=27.6~\mu$m; (b)
  $d=30.4~\mu$m; and (c) $d = 33.2~\mu$m. The spectral profiles and the
  cavity reflection profile (dashed green lines) are centered at the
  same energy $E_{\ind{c}}=6.9558$~keV.  The x-rays are at the
  incidence angle $\theta=25.6434^{\circ}$ to the diffraction planes
  (111) and at $88^{\circ}$ to the (400) planes in the cavity
  crystals.}
\label{fig004}
\end{figure}


Intensity oscillations -- equal thickness fringes -- are observed on the tails of the reflection
dependence.  The period of the oscillations is  
\begin{equation}
\delta E\, =\,
\frac{hc}{2d}\,\frac{\gamma_{\ind{H}}}{\sin^2\theta}\hspace{0.25cm}  \mathrm{or} \hspace{0.25cm} \delta E\, =\ \frac{hc}{2d\sin\theta}\hspace{0.25cm}  \mathrm{if} \hspace{0.25cm} \eta=0,
\label{eq0030}
\end{equation}
see Eq.~(2.176) of \cite{Shvydko-SB}. In the example shown in
Fig.~\ref{fig003}, $\delta E=95$~meV for a $d$=15-$\mu$m-thick crystal
at $\theta=25.64^{\circ}$.

For a much thicker crystal, the period becomes very small and the
oscillations wash out, as illustrated by the dashed green line in
Fig.~\ref{fig003} calculated for  $d$=300-$\mu$m-thick cavity crystals
A, B, C, and D in the (400) Bragg refection with a spectral width of
$\Delta E_{\ind{400}}=54$~meV.

If the period of oscillations is tailored to be larger than the cavity
bandwidth, the fringes can be tuned to the center $E_{\ind{c}}$ of the
cavity band with a purpose of outcoupling a desired amount of the
intracavity power. The tuning is achieved by varying the glancing
angle of incidence $\theta$ of x-rays to the reflecting atomic
planes. In the example shown in Fig.~\ref{fig002}, $\delta E=95$~meV,
while the cavity bandwidth $\Delta E_{\ind{400}}=54$~meV. The amount
of the intracavity power, which can be outcoupled, changes from almost
zero to $\simeq$~30-40\%, i.e., much larger than what the permeable
thin-crystal outcoupling technique can realistically provide.

\begin{figure*}[t!]
\includegraphics[width=0.99\textwidth]{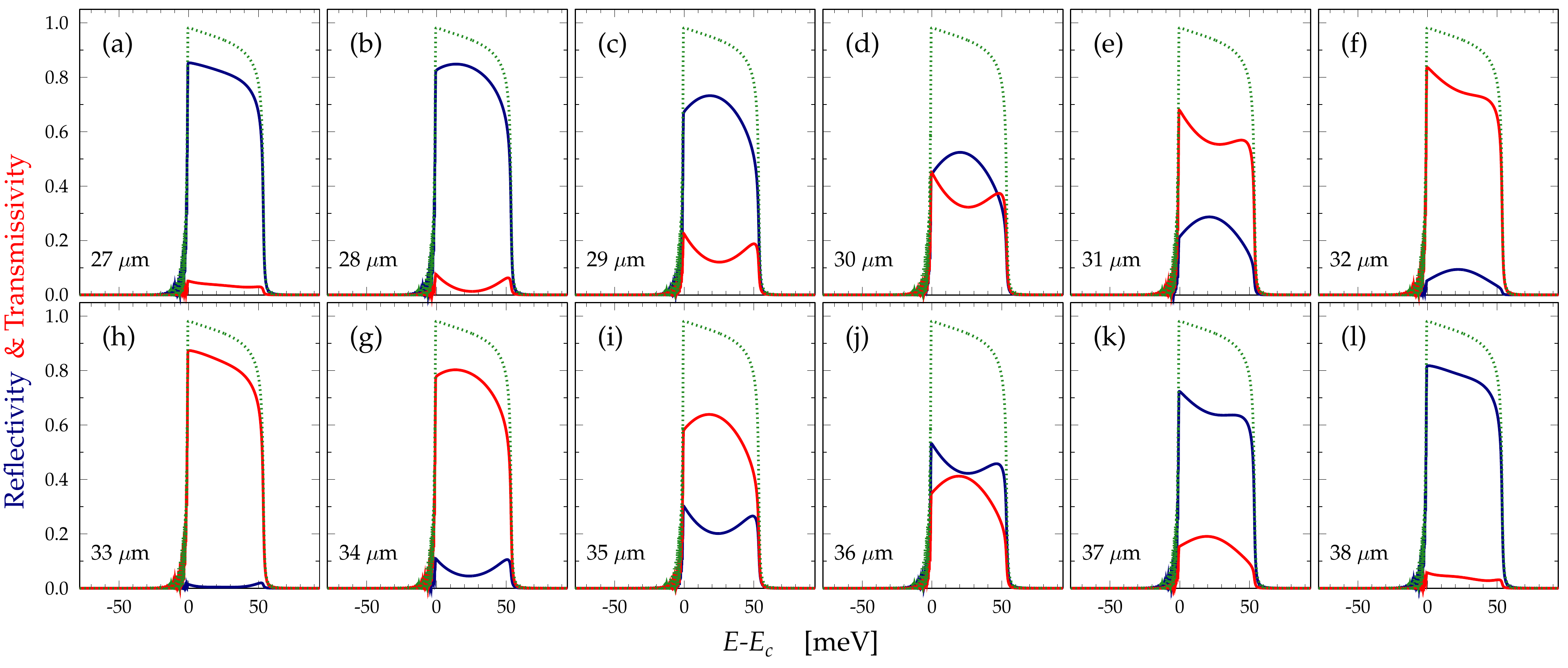}
\caption{Similar to Fig.~\ref{fig002}; however, the beam-splitter
  crystal is in the transmission (Laue-case) geometry; see
  Fig.~\ref{fig001LB}(b). The beam-splitter crystal thickness $d$
  varies from $27~\mu$m (a) to $38~\mu$m (l). Other parameters are
  provided in the caption to Fig.~\ref{fig004}.  Instead of varying
  the crystal thickness, azimuthal angle $\phi$ can be varied
  alternatively by crystal 
rotation about the diffraction vector
in the range from about 0 to $35^{\circ}$.}
\label{fig006}
\end{figure*}


\subsection{Laue-case beam splitter}

Even larger outcoupling efficiency can be achieved with the
Bragg-diffracting beam-splitter crystal in the transmission
(Laue-case) scattering geometry as in Fig.~\ref{fig001LB}(b). This is
possible due to the Pendell{\"o}sung effect \cite{Ewald17}, a unique
feature for this scattering geometry; see also \cite{Authier}.  In the
Laue-case scattering geometry, both the Bragg-diffracted (BD) and
forward-Bragg-diffracted (FBD) beams are on the same side of the
crystal.  There is a periodic exchange of energy between the two
beams, propagating through the crystal thickness, as between two
coupled pendulums. When one is in maximum, another is in minimum and
vice versa. As a result, the BD and FBD intensities are complementary
oscillating functions of the crystal thickness, the Pendell{\"o}sung
effect, which is illustrated in Figs.~\ref{fig004}(a)-(c) on the
example of the spectral profiles calculated for three different
crystal thicknesses $d$ at a fixed glancing angle of incidence
$\theta$.  There is almost zero transmission and $\simeq 90$\%
reflectivity at $E=E_{\indrm{c}}$ in Fig.~\ref{fig004}(a), while the
picture reverses in Fig.~\ref{fig004}(c).  The Pendell\"osung period
is equal to $\Lambda_{\ind{H}} =2\pi\bar{\Lambda}_{\ind{H}}$, where
$\bar{\Lambda}_{\ind{H}}$ is the extinction length given by
Eq.~\eqref{eq0010}. In particular, for the case of the beam-splitter
crystal presented in Fig.~\ref{fig004},
  $\bar{\Lambda}_{\ind{111}}^{(s)}=1.1~\mu$m,
  $\bar{\Lambda}_{\ind{111}}=1.75~\mu$m, and $\Lambda_{\ind{111}}=11~\mu$m. 

  Furthermore, the BD and FBD intensities are also complementary
  oscillating functions of the photon energy, featuring the equal
  thickness fringes as in the Bragg-case geometry; see
  Fig.~\ref{fig003}.

  The width of the central maxim (minimum) at $E=E_{\ind{c}}$ in the
  Laue-case diffraction is roughly about two periods of the equal
  thickness fringes, given by Eq.~\eqref{eq0030}, and amounts to
  $\simeq 120$~meV in the present case.  Because the width is much
  larger than the 54-meV bandwidth of the cavity, this beam splitter
  can outcouple efficiently the intracavity power, as the results of
  calculations show in Fig.~\ref{fig006}.  When the crystal thickness
  is $d=2.5 \Lambda_{\ind{H}} \simeq 27~\mu$m, the maximum
  reflectivity is achieved and almost 90\% of the intracavity power is
  outcoupled; see Fig.~\ref{fig006}(a). If $d=3 \Lambda_{\ind{H}}
  \simeq 33~\mu$m, the reflectivity is lowest and the transmissivity
  is highest, as in Fig.~\ref{fig006}(h). Figures~\ref{fig006}(h)-(l)
  show the reverse process. Crystal thickness variation can be
  accomplished, for example, if the crystal has a wedge
  form. Alternatively, instead of physically changing the crystal
  thickness in this range, the azimuthal angle can be varied in the
  range from about 0 to $35^{\circ}$ by rotating the crystal around
  the (111) diffraction vector.

\begin{figure*}[t!]
\includegraphics[width=0.99\textwidth]{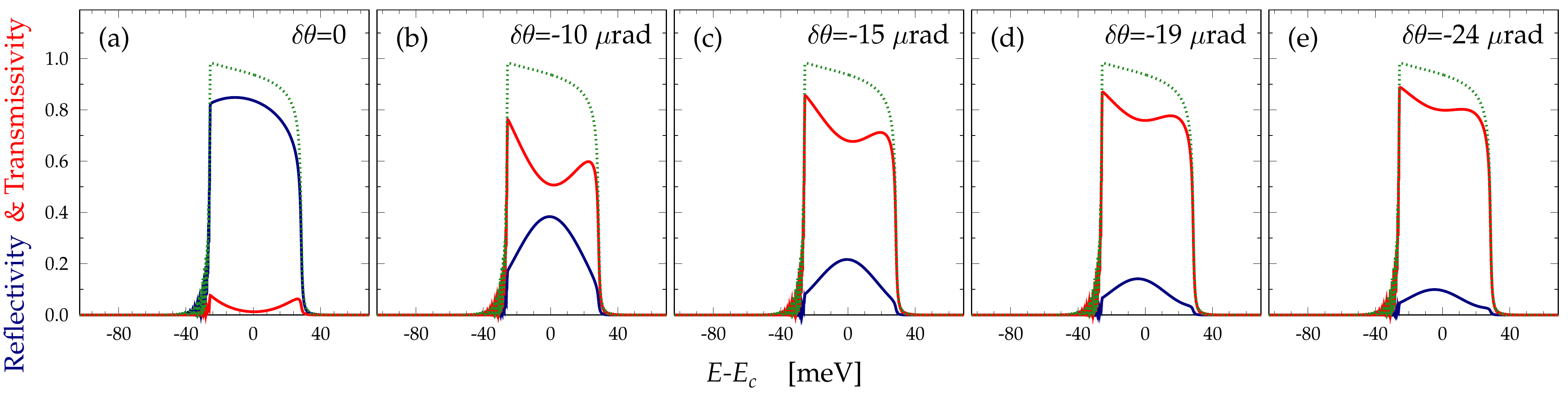}
\caption{Similar to Fig.~\ref{fig006}, however the beam splitter
shown in  Fig.~\ref{fig001LB}(b)
is a diamond crystal with a fixed 28-$\mu$m thickness  set in the
  (111) Bragg reflection at incidence angles
  $\theta=\tilde{\theta}+\delta\theta$
  ($\tilde{\theta}=25.6434^{\circ}$) as in Fig.~\ref{fig004}(a).}
\label{fig007}
\end{figure*}

If the width of the equal thickness fringes are broader or equal to
the cavity bandwidth, outcoupling can also be achieved by tuning the
fringes to the cavity bandwidth center by properly selecting the
glancing angle of incidence $\theta$, as the results of calculations
show in Fig.~\ref{fig007}. Varying the outcoupling efficiency by
changing the angle is straightforward; however, with this approach it
is difficult to reach the 90\% outcoupling efficiency possible by
crystal thickness variation and the Pendell\"osung effect, as in
Fig.~\ref{fig006}.  These results are similar to outcoupling using
equal thickness fringes in the Bragg-case geometry presented in
Fig.~\ref{fig002}; however, in the Laue-case, the crystal thickness can
be several times larger, which is an advantage.

Altogether, outcoupling with the beam splitter in the Laue-case
geometry seems to be preferred compared to the beam-splitter case in
the Bragg-case geometry.  However, there is a complication in the
Laue-case, which has to be handled appropriately. If the asymmetry
angle $\eta$ is nonzero, as in the Laue-case geometry (see
Fig.~\ref{fig004}) angular dispersion takes place, resulting in
detrimental distortions of the wavefront and coherence; see, for
example, \cite{BSS95,SDSS99,Shvydko-SB,SL12}.  To undo this effect,
Bragg diffraction from the second crystal S$_{\ind{2}}$ can be applied
identical to S$_{\ind{1}}$, as in Fig.~\ref{fig001LB}(b); however, in
a time-reversed setting with the asymmetry angle $\pi-\eta$, where
$\eta$ is the asymmetry angle of the first crystal. Crystal
S$_{\ind{2}}$ should have a thickness of $(n+1/2)\Lambda_{\ind{H}}$,
where $n=0,1,2 ,..$ ensuring the highest reflectivity in Laue-case
geometry.

\section{Discussion}

In the present paper we consider different approaches for coupling
x-rays out of XFEL cavities. A standard approach of using permeable
thin-crystal Bragg reflecting mirrors is very often limited to
extraction only a few percentage points of the intracavity's
power. This is acceptable for low-gain XFEL oscillators. The high-gain
regenerative amplifier XFEL requires, however, much higher outcoupling
efficiency. Using Bragg reflecting pinhole crystal mirrors for this
purpose is a possibility discussed in \cite{FSS19}.  Here we analyze
an alternative approach: intracavity Bragg-reflecting crystal 
beam splitters.

There are significant advantages in this approach. First, the
out-coupled power can be varied over a large range, from almost zero
to nearly 100\%, by varying the crystal thickness or
orientation. Second, extremely thin crystals are not required. The
crystals have only to be much thinner than the absorption length in
diamond. Third, fundamental and higher harmonics can be out-coupled at
different locations. Fourth, multi-beam out-coupling to increase the
number of users can be achieved by installing additional beam-splitter
crystal pairs $S_{\ind{1}}$-$S_{\ind{2}}$ at different locations in
the cavity.

Other types of beam splitters can be used as well.  Grazing incidence
mirrors or crystals in Bragg diffraction with sharp edges can be used
as wavefront-division beam splitters \cite{HOM18}. X-ray transparent
diamond diffraction gratings also can be considered
\cite{KRM12,MKZ15,KLi2019}. However, there are always two beams in
each diffraction order, and several diffraction orders can
contribute. Many beams could be good or bad depending on the
application.\\

\section{Acknowledgments}

Kwang-Je Kim (Argonne National Laboratory) is acknowledged for the
discussion of importance of efficient outcoupling schemes.  Anne
Sakdinawat, David Attwood, and Diling Zhu (SLAC National Laboratory)
are acknowledged for sharing results on  hard-x-ray high-fidelity diamond-grating
beam splitters.  Work at Argonne National Laboratory
was supported by the U.S. Department of Energy, Office of Science,
Office of Basic Energy Sciences, under contract DE-AC02- 06CH11357.


\end{document}